\documentclass[aps,reprint,superscriptaddress,nofootinbib,showpacs]{revtex4-1}
\usepackage{amsmath,latexsym,amssymb,hyperref,graphicx}
\usepackage{slashed}
\usepackage{enumerate}
\usepackage{subfigure}
\usepackage{float}

\hypersetup{colorlinks,citecolor= nicegreen,linkcolor= niceblue}
\usepackage{color}
\definecolor{nicered}{rgb}{0.7,0.1,0.1}
\definecolor{nicegreen}{rgb}{0.1,0.5,0.1}
\definecolor{niceblue}{rgb}{0.1,0.1,0.8}

\begin{document}

\title{$SO(10)$ thick branes and perturbative stability}

\author{Rafael Chavez}
\affiliation{Departamento de Ciencias B\'asicas, Universidad Polit\'ecnica Salesiana, Ecuador}

\author{Rommel Guerrero}
\affiliation{Departamento de F\'isica, Universidad Centroccidental Lisandro Alvarado, Venezuela}

\author{R. Omar Rodriguez}
\affiliation{Departamento de F\'isica, Universidad Centroccidental Lisandro Alvarado, Venezuela}

\begin{abstract}
 \noindent
Three self-gravitating $SO(10)$ domain walls in five dimensions are obtained and their properties are analyzed. These non-abelian domain walls interpolate between  AdS${}_5$ spacetimes  with different embedding of $SU(5)$ in $SO(10)$ and they can be distinguished, among other features, by the unbroken group on each wall, being either $SO(10)$, $SO(6)\otimes SU(2)\otimes U(1)/Z_2$  or $SU(4)\otimes SO(2)\otimes U(1)/Z_4$. We show that, unlike  Minkowskian versions, the curved scenarios are perturbatively  stable due to the gravitational capture of scalar fluctuations associated to the residual orthogonal subgroup in the core of the walls.  These stabilizer modes are additional to the four-dimensional Nambu-Goldstone states found in two of the three gravitational sceanarios.
\end{abstract}

\maketitle

\section{Introduction}

Our universe could be a hypersurface embedded in a higher dimensional spacetime and among the proposals that have emerged to develop this idea, the five-dimensional Randall-Sundrum model \cite{Randall:1999vf} has received much attention because standard gravitation can be recovered on the four-dimensional worldsheet (or 3-brane) of the scenario. For a discussion about localization of matter and interaction fields, see \cite{Bajc:1999mh,Dvali:2000rx}.

In more realistic models the thickness of the worldsheet is taken into account, in this case the brane is generated by a domain wall, a solution to Einstein gravity theory interacting with a scalar field where the scalar field is a standard kink interpolating between the minima of a potential with spontaneously broken symmetry \cite{Gremm:1999pj,DeWolfe:1999cp,Wang:2002pka,Bazeia:2003aw, Melfo:2002wd,CastilloFelisola:2004eg,Melfo:2006hh,Guerrero:2006gj,Guerrero:2009ac}. This scenarios are topologically stables and, consequently, the analysis of small fluctuations of both the metric tensor and the scalar field \cite{Giovannini:2001fh},  revels  a tower of modes free of tachyonic instabilities.

Domain walls generated from several scalar fields have been also considered, see Ref. \cite{George:2011tn}, and among other properties, it is observed that the flat configuration admits the translation zero mode in  Kaluza-Klein spectrum of the scalar perturbations  which is removed when the extra dimension is warped; however,  the general set-up, in the presence of gravity, can support one or several extra zero modes in the excitations tower.

It is also possible to consider domain walls with multiple scalar fields in terms of a non-abelian  source with internal gauge symmetry group $G$, which is advantageous on the wall, where our universe is realized,  because a symmetry breaking pattern, $G\rightarrow H_0$, could be obtained. This opens up the possibility of building braneworld with Standard Model group on the wall. In this sense several attempts  have been made; in particular, a pair of perturbatively stable self-gravitating $SU(5)\otimes Z_2$ domain walls, with different group $H_0$, were reported in \cite{Melfo:2011ev}. Remarkably, one of them corresponds to curved version of the flat solution found in \cite{Pogosian:2000xv} and widely disscused in \cite{Vachaspati:2001pw,Pogosian:2001pq,Vachaspati:2003zp}. Other notable attempt, with $G=E_6$ but in flat space, was reported in \cite{Davidson:2007cf}.

The perturbative stability analysis of the  $SU(5)\otimes Z_2$ walls  was performed in \cite{Pantoja:2015yin} (as far as we know, there is no a topologiacal charge difined for non-abelian walls) and, in addition to verifying the local stability of scenarios, it was shown  that, for a four-dimensional observer localized on the brane,  the tensor and vector sectors of the gravity fluctuations behave  in a similar way to the  abelian domain wall set-up \cite{Giovannini:2001fh}; namely, while the zero mode of the tensor excitations is localized, there is not a nomalizable solution for the vector perturbations.
On the other hand, in the spectrum of the scalar fluctuations, the absence of the translation mode was verified
and normalizable massless scalar modes associated to the particular symmetry breaking pattern considered on the wall were found.

From the point of view of Grand Unified Theories, the symmetry $O(10)$ is considered more fundamental that $U(5)$ in the sense that $SU(5)\subset SO(10)$ and the Standard Model group is embedded in $SO(10)$ as a single irreducible representation of the underlying gauge group. In \cite{Shin:2003xy}, three flat $SO(10)$ domain walls were found and, just as in $SU(5)\otimes Z_2$ case, a symmetry breaking pattern, $SO(10)\rightarrow H_0$, was determined for each wall. These scenarios will be considered in this paper; concretely,  we will focus on both the extension to curved spacetime  and the stability under small perturbations. Among the results that we will show highlight, the local instability of two of the flat scenarios due to tachyonic P\"oschl-Teller modes in spectrum of scalar perturbations, which, fortunately,  can be removed when gravity is included; and, the four-dimensional localization of massless scalar states along the broken generators associated to $SO(10)\rightarrow H_0$, which occurs only when gravity is present in the model.

The paper is organized as follows, in  Sections \ref{solutions1} and \ref{solutions} the gravity $SO(10)$ set-up and the extensions to warped spacetime of the flat $SO(10)$ kinks are obtained.
In Section \ref{excitations}, the perturbative stability analysis of the $SO(10)$ walls is performed and it is show that gravitation rescues the stability  through the capture of massless scalar perturbations associated to the orthogonal subgroup of $H_0$.
Finally, in Section \ref{summary} our  results and conclusions are summarized  and  presented.

\section{Self-gravitating $SO(10)$ kinks}\label{solutions1}

Consider the Einstein-scalar field coupled system in five dimensions
\begin{eqnarray}\label{Einstein}
&& R_{ab}-\frac{1}{2}g_{ab}R =\nonumber\\ -\frac{1}{2}\text{Tr}(\nabla_a\mathbf{\Phi}\nabla_b\mathbf{\Phi})&+&g_{ab}(\frac{1}{4}\text{Tr}(\nabla_c\mathbf{\Phi}\nabla^c\mathbf{\Phi})-V(\mathbf{\Phi}))
\end{eqnarray}
and
\begin{equation}\label{eqfield}
\nabla_c\nabla^c\mathbf{\Phi}=\frac{\partial V(\mathbf{\Phi})}{\partial\mathbf{\Phi}},
\end{equation}
where $\mathbf{\Phi}$ is a scalar multiplet in the 45-adjoint representation of $SO(10)$, i.e
\begin{equation}\label{transformation}
\mathbf{\Phi}\rightarrow{\boldsymbol{O}}\mathbf{\Phi}{\boldsymbol{O}}^\text{T}, \quad
{\boldsymbol{O}}=e^{\frac{1}{2}\alpha_{j_1 j_2}\mathbf{L}_{j_1 j_2}}
\end{equation}
with  $\alpha$ and $\mathbf{L}$ the parameters and generators of the group respectively. In particular, for the generators in the fundamental representation we have
\begin{equation}
\left(\mathbf{L}_{j_1 j_2}\right)_{j_3 j_4}=\delta_{j_1 j_4}\delta_{j_2 j_3}-\delta_{j_1 j_3}\delta_{j_2 j_4}.
\end{equation}
The latin index $j=1, \dots , 10$, denotes an internal index of SO(10) group.

Now, consider the spacetime ($\mathbb{R}^5$, $\bf{g}$) where the tensor metric $\bf{g}$ for a five-dimensional static spacetime with a
planar-parallel symmetry, in a particular coordinate basis, is given by
\begin{equation}\label{metric}
ds^2=e^{2A(y)}\eta_{\mu\nu}dx^\mu dx^\nu + dy^2,\quad \mu,\nu=0,\dots, 4.
\end{equation}
We are interested in the realization of brane worlds on this geometry. In order to do this, we consider a potential of sixth-order
\begin{eqnarray}\label{potencialSO(10)}
V(\mathbf{\Phi})&=&V_0+\frac{\mu^2}{2}\text{Tr}\mathbf{\Phi}^2+\frac{h}{4}(\text{Tr}\mathbf{\Phi}^2)^2+\frac{\lambda}{4}\text{Tr}\mathbf{\Phi}^4
\nonumber\\&&+\frac{\alpha}{6}(\text{Tr}\mathbf{\Phi}^2)^3
+\frac{\gamma}{6}\text{Tr}\mathbf{\Phi}^4\text{Tr}\mathbf{\Phi}^2+\frac{\beta}{6}\text{Tr}\mathbf{\Phi}^6,
\end{eqnarray}
where $V_0$ is a constant to be fixed.

Two aspects of the theory should be highlighted. First, the $O(10)$ group has two disconnected components, the $SO(10)$ special subgroup and the antispecial part. These two subspaces are related by a discrete  $Z_2$ transformation of $O(10)$. Second, both $\text{Tr}(\mathbf{\Phi}^3)$ and $\text{Tr}(\mathbf{\Phi}^5)$  vanishes in (\ref{potencialSO(10)}) because the scalar field is antisymmetric. Therefore, the reflection symmetry
\begin{equation}
Z_2: \mathbf{\Phi}\rightarrow -\mathbf{\Phi},
\end{equation}
that connects the vacuum expectation values (vev's) of scalar field
\begin{equation}
\mathbf{\Phi}(y=-\infty)=-{\boldsymbol{O}} \mathbf{\Phi}(y=+\infty){\boldsymbol{O}}^\text{T} ,
\end{equation}
is part of the model and is outside the $SO(10)$ group \cite{Shin:2003xy}. Hence, a $SO(10)$ kink  interpolating between two minima of $V(\mathbf{\Phi})$ is a feasible solution for the coupled system. A kink solution for a sixth-order polynomial potential and a single self-gravitating scalar field have been found in \cite{DeWolfe:1999cp}.

We will assume that the scalar field takes values in the Cartan-subalgebra space of $SO(10)$, that is
\begin{equation}
\mathbf{\Phi}=\phi_1\mathbf{L}_{12}+\phi_2\mathbf{L}_{34}+\phi_3\mathbf{L}_{56}+\phi_4\mathbf{L}_{78}+\phi_5\mathbf{L}_{90}.
\end{equation}
(Hereon we denote the subscript $10$ as $0$.) Following the usual strategy, from  (\ref{Einstein}) and (\ref{eqfield}) we find
\begin{eqnarray}
3A^{\prime\prime}=-\phi_k^\prime\phi_k^\prime\ ,\quad \frac{3}{2}A^{\prime\prime}+6A^{\prime2}=-V({\bf\Phi})
\end{eqnarray}
and
\begin{eqnarray}\label{eccSO(10)}
\phi_i^{\prime\prime}+4A^\prime \phi_i^\prime=-2(\mu^2 &-& 2 h \phi_k\phi_k)\phi_i+2\lambda \phi_i^3+2\beta \phi_i^5
\nonumber\\+\frac{4\gamma}{3} (2\phi_i^2\phi_k\phi_k&-&\phi_k^2\phi_k^2) \phi_i-8\alpha(\phi_k\phi_k)^2 \phi_i ,
\end{eqnarray}
where prime indicates derivative with respect to extra coordinate $y$ and $i, k=1,\dots, 5$.

From the minimum of the potential we take the following three boundary conditions \cite{Li:1973mq, Shin:2003xy}
\begin{equation}\label{vacioA}
{\mathbf{\Phi_A}}(y=\pm\infty)=\pm \frac{v}{\sqrt{5}}\ (\mathbf{L}_{12}+\mathbf{L}_{34}+\mathbf{L}_{56}+\mathbf{L}_{78}+\mathbf{L}_{90}),
\end{equation}
\begin{equation}\label{vacioB}
{\mathbf{\Phi_B}}(y=\pm\infty)=\frac{v}{\sqrt{3}}\ (\pm \mathbf{L}_{12}\pm \mathbf{L}_{34}\pm \mathbf{L}_{56}-\mathbf{L}_{78}-\mathbf{L}_{90})
\end{equation}
and
\begin{equation}\label{vacioC}
{\mathbf{\Phi_C}}(y=\pm\infty)= v\ (\pm\mathbf{L}_{12}-\mathbf{L}_{34}-\mathbf{L}_{56}-\mathbf{L}_{78}-\mathbf{L}_{90}).
\end{equation}

Now, choosing 
\begin{equation}
h=0,\quad \alpha=0, \quad \gamma=0,
\end{equation}
in order to decouple (\ref{eccSO(10)}), and 
solving the boundary value problem for
\begin{equation}\label{warp}
A=-\frac{v^2}{9}\left[2\ln\cosh(k y)+\frac{1}{2}\tanh^2(k y)\right],
\end{equation}
we obtain three non-abelian kink solutions determined as follows

 {\it Symmetric kink}: for the boundary condition ({\ref{vacioA}}), we get a kink solution with a single component,
\begin{equation}
\mathbf{\Phi_A}=v\tanh(ky)\mathbf{M_A}, \label{a}
\end{equation}
such that
\begin{equation}
\mathbf{M_A}=\frac{1}{\sqrt{5}}(\mathbf{L}_{12}+\mathbf{L}_{34}+\mathbf{L}_{56}+\mathbf{L}_{78}+\mathbf{L}_{90})
\end{equation}
and 
\begin{eqnarray}
v=\frac{\sqrt{5}}{2}&&\sqrt{\frac{\lambda}{\beta}-\frac{9}{10}},\quad k=\frac{3}{2\sqrt{10}}\sqrt{\beta\left(\frac{\lambda}{\beta}-\frac{9}{10}\right)},\\
&&\mu=\frac{\sqrt{3}}{4}\sqrt{\beta\left(\frac{\lambda}{\beta}+\frac{3}{10}\right)\left(\frac{\lambda}{\beta}-\frac{9}{10}\right)} ,\\
&&\qquad V_0=\frac{9}{64}\beta\left(\frac{\lambda}{\beta}-\frac{9}{10}\right)^2 ,
\end{eqnarray}
with $\beta>0$ and $\lambda>9\beta/10$.

{\it Asymmetric kink}: in connection with (\ref{vacioB}),  the kink solution obtained in this case has two components,
\begin{equation}
\mathbf{\Phi_B}=v\tanh(ky)\mathbf{M_B}-\sqrt{\frac{{2}}{3}}v\mathbf{P_B},\label{b}
\end{equation}
where
\begin{equation}
\mathbf{M_B}=\frac{1}{\sqrt{3}}(\mathbf{L}_{12}+\mathbf{L}_{34}+\mathbf{L}_{56}),\ \mathbf{P_B}=\frac{1}{\sqrt{2}}(\mathbf{L}_{78}+\mathbf{L}_{90})
\end{equation}
and 
\begin{eqnarray}
v=&&\frac{\sqrt{3}}{2}\sqrt{\frac{\lambda}{\beta}-\frac{3}{2}},\quad k=\frac{\sqrt{3}}{2\sqrt{2}}\sqrt{\beta\left(\frac{\lambda}{\beta}-\frac{3}{2}\right)},\\
&&\mu=\frac{\sqrt{3}}{4}\sqrt{\beta\left(\frac{\lambda}{\beta}+\frac{1}{2}\right)\left(\frac{\lambda}{\beta}-\frac{3}{2}\right)} ,\\
&&V_0=\frac{1}{24}\beta \left(\frac{\lambda}{\beta}+\frac{33}{8}\right)\left(\frac{\lambda}{\beta}-\frac{3}{2}\right)^2 ,
\end{eqnarray}
with $\beta>0$ and $\lambda>3\beta/2$.

{\it Superasymmetric kink}: for the  condition (\ref{vacioC}), as in the previous case, we find a kink solution with components in two directions,
\begin{equation}
\mathbf{\Phi_C}=v\tanh(ky)\mathbf{M_C}-2v\mathbf{P_C},\label{c}
\end{equation}
where
\begin{equation}
\mathbf{M_C}=\mathbf{L}_{12},\quad\mathbf{P_C}=\frac{1}{2}(\mathbf{L}_{34}+\mathbf{L}_{56}+\mathbf{L}_{78}+\mathbf{L}_{90}).
\end{equation}
and
\begin{eqnarray}
v=&&\frac{1}{2}\sqrt{\frac{\lambda}{\beta}-\frac{9}{2}},\quad k=\frac{3}{2\sqrt{2}}\sqrt{\beta\left(\frac{\lambda}{\beta}-\frac{9}{2}\right)},\\
&&\mu=\frac{\sqrt{3}}{4}\sqrt{\beta\left(\frac{\lambda}{\beta}+\frac{3}{2}\right)\left(\frac{\lambda}{\beta}-\frac{9}{2}\right)} ,
\end{eqnarray}
\begin{equation}
V_0=\frac{1}{12}\beta \left(\frac{\lambda}{\beta}+\frac{63}{16}\right)\left(\frac{\lambda}{\beta}-\frac{9}{2}\right)^2 ,
\end{equation}
such that  $\beta>0$ and $\lambda>9\beta/2$.

In all cases $\mathbf{M_{A,B,C}}$ and $\mathbf{P_{B,C}}$ are orthogonal generators of $SO(10)$.

The warp factor  (\ref{warp}) together with (\ref{a}), (\ref{b}) or (\ref{c}) represent a two-parameter family of $SO(10)$ static domain walls, asymptotically AdS${}_5$ with cosmological constant determined by
\begin{equation}
\Lambda_{\bf A}=-\frac{5}{48}\beta\left(\frac{\lambda}{\beta}-\frac{9}{10}\right)^3
\end{equation}
for the symmetric case;
\begin{equation}
\Lambda_{\bf B}=-\frac{1}{16}\beta\left(\frac{\lambda}{\beta}-\frac{3}{2}\right)^3
\end{equation}
for the asymmetric case; or
\begin{equation}
\Lambda_{\bf C}=-\frac{1}{48}\beta\left(\frac{\lambda}{\beta}-\frac{9}{2}\right)^3
\end{equation}
for the superasymmetric case. On the other hand, they can also be considered as the extensions to curved spacetime of the flat $SO(10)$ kinks reported in  \cite{Shin:2003xy} and supported on a four-order potential ( $\alpha=\gamma=\beta=0$). In this case the system is decoupled for $h=0$ and is satisfied when $k=\mu$ and $v=\sqrt{5\mu^2/\lambda}$, $\sqrt{3\mu^2/\lambda}$, $\mu/\sqrt{\lambda}$ respectively for the symmetric,  asymmetric and superasymmetric kink.

\section{The breaking scheme of $SO(10)$ Brane}\label{solutions}

Any of the non-abelian kinks induces the breaking of $SO(10)$ both in the core and at the edge of the scenarios. The unbroken symmetry at $y\rightarrow\pm\infty$, for each kink solution, is given by
\begin{equation}
SO(10)\longrightarrow \frac{SU(5)\otimes U(1)}{Z_5},
\end{equation}
and in concordance with the boundary conditions  (\ref{vacioA}, \ref{vacioB}, \ref{vacioC}), $SU(5)$ is embedded in $SO(10)$ in different ways \cite{Li:1973mq}.

In the core of the wall, the remaining groups are completely different. For the symmetric kink (\ref{a}) the scalar field vanish in $y=0$; so, all generators of $SO(10)$ are annihilated for the field and the group is preserved on the wall. This is a straightforward generalization of the abelian case.

For the other scenarios the situation is more interesting. This means that some components of ${\mathbf{\Phi_{B,C}}}$ can be nonzero in the core, and some generators of $SO(10)$ remain broken even in the core. Therefore, the spontaneous symmetry breaking is non-trivially realized on the wall. To see this explicitly we consider a combinations $\mathbf{T}$ of generators $\mathbf{L}$ such that $\partial^2V({\mathbf{\Phi}})/\partial\phi_{j_1}\partial\phi_{j_2}$ is diagonal for each kink solution $\mathbf{\Phi_{B,C}}$. For the asymmetric kink we find that  $SO(3)$ sector of  $SO(10)$ is isomorphically equivalent to $SU(2)$; on the other hand, for the superasymmetric kink, we get that $SO(6)$ sector of $SO(10)$ becomes isomorphic to $SU(4)$.

Therefore, in the core of the non-abelian walls (\ref{b}) and (\ref{c}) respectively we have
\begin{equation}\label{s2}
SO(10)\longrightarrow H_\mathbf{B}=\frac{SO(6)\otimes SU(2)\otimes U(1)_{\mathbf{P_B}}}{Z_2}
\end{equation}
and
\begin{equation}\label{s3}
SO(10)\longrightarrow H_{\mathbf{C}}=\frac{SU(4)\otimes SO(2)\otimes U(1)_{\mathbf{P_C}}}{Z_4}.
\end{equation}

We leave in the Appendix the technical details associated to (\ref{s2}) and (\ref{s3}).

\section{Stability of non-abelian kink}\label{excitations}

These non-abelian walls are not topologically protected and, therefore, their stability is not guaranteed. Let us examine the perturbative stability of these domain wall spacetimes considering small deviations to the solutions of the Einstein scalar field equations, $g_{ab}$ and $\mathbf{\Phi}$, defined by $h_{ab}$ and $\boldsymbol{\varphi}$, respectively.

Thus, in accordance with Ref. \cite{CastilloFelisola:2004eg}, from (\ref{Einstein}) and (\ref{eqfield})  the equations for the excitations are obtained
\begin{eqnarray}\label{gravfluctuation}
-\frac{1}{2}g^{cd}\nabla_c\nabla_d h_{ab}&+&R^c{}_{(ab)}{}^dh_{cd}+R^c{}_{(a}h_{b)c}\nonumber\\-\frac{1}{2}\nabla_a\nabla_b\left(g^{cd}h_{cd}\right)&+&\nabla_{(a}\nabla^c h_{b)c}=
 2\nabla_{(a}\phi_{j}\nabla_{b)}\varphi_{j}\nonumber\\+\frac{2}{3}h_{ab}V(\mathbf{\Phi})&+&\frac{2}{3}g_{ab} \frac{\partial V(\mathbf{\Phi})}{\partial\phi_{j}}\varphi_{j}
\end{eqnarray}
and
\begin{eqnarray}\label{scalarfluctuation}
-h^{ab}\nabla_a\nabla_b\phi_{j_1}&-&\frac{1}{2}g^{ab}g^{cd}\left(\nabla_ah_{bd}+\nabla_bh_{ad}-\nabla_dh_{ab}\right)\nabla_c\phi_{j_1}\nonumber
\\&+&g^{ab}\nabla_a\nabla_b\varphi_{j_1}=\frac{\partial^2 V(\mathbf{\Phi})}{\partial\phi_{j_2}\partial\phi_{j_1}}\varphi_{j_2}
\end{eqnarray}
where we have considered that $\mathbf{\Phi}=\phi_j\mathbf{T}^j$ and $\boldsymbol{\varphi}=\varphi_j\mathbf{T}^j$.

Now, taking into account the generalization of the Bardeen formalism \cite{Bardeen:1980kt} to the case of warped geometries presented in \cite{Giovannini:2001fh}, we consider  the decomposition of $h_{ab}$ in terms of  tensor, vector and scalar modes, namely
\begin{eqnarray}
h_{\mu\nu}=2e^{2A}(h_{\mu\nu}^{TT}+\partial_{(\mu}f_{\nu)}&+&\eta_{\mu\nu}\psi +\partial_\mu\partial_\nu E),\label{hmn}\\
h_{\mu y}=e^{A}\left(D_\mu+\partial_\mu C\right) &,&  h_{yy}=2\omega,\label{hyy}
\end{eqnarray}
In order to preserve the degrees of freedom of $h_{ab}$, both $h^{TT}_{\mu\nu}$ and $f_\mu$ and $D_\mu$ must satisfy the conditions of transverse traceless and divergence-free
\begin{eqnarray}
h^{TT\mu}_\mu=0, \ \partial^\mu h^{TT}_{\mu\nu}=0,\ \partial^\mu f_\mu=0,\ \partial^\mu D_\mu=0,
\end{eqnarray}

These modes can be rewritten in terms of following variables: a vector field
\begin{equation}
V_\mu=D_\mu-e^{A}f^\prime_\mu,\label{V}
\end{equation}
two scalar fields
\begin{eqnarray}
\Gamma=\psi-&&A^\prime(e^{2A}E^\prime- e^AC),\\
\Theta=\omega &&+(e^{2A}E^\prime-e^AC)^\prime.\label{Theta}
\end{eqnarray}
and the non-abelian scalar
\begin{equation}
\boldsymbol\chi=\boldsymbol\varphi-\mathbf{\Phi}^\prime\left(e^{2A}E^\prime-e^AC\right);\label{chi}
\end{equation}
which, similarly to $h^{TT}_{\mu\nu}$,  do not change under the following infinitesimal coordinate transformation 
\begin{equation}\label{infinitesimal}
x^a\rightarrow\bar{x}^a=x^a+\epsilon^a,
\end{equation}
with
\begin{equation}\label{parameters}
\epsilon_a=(e^{2A}\epsilon_\mu, \epsilon_y),
\end{equation}
\begin{equation}
\epsilon_\mu=\partial_\mu\epsilon+\zeta_\mu,\quad \partial^\mu\zeta_\mu=0 .
\end{equation}

Choosing the longitudinal gauge, $E=0$, $C=0 $ and $f_\mu=0$, from (\ref{gravfluctuation}) and (\ref{scalarfluctuation}) the equations for the gauge-invariant variables are obtained which we write below in conformal coordinates,  $dz=\exp(-A(y))dy$.

{\it Graviton and graviphoton}: whilenamely the tensor modes equation is determined by
\begin{equation}\label{graviton}
\left(-\partial_z^2 +V_Q\right)\psi_{\mu\nu}(z)=m^2\psi_{\mu\nu}(z),
\end{equation}
where $\psi_{\mu\nu}\equiv e^{3A/2}h_{\mu\nu}$ and
\begin{equation}
V_Q=\frac{9}{4}A^{\prime 2}+\frac{3}{2}A^{\prime\prime};
\end{equation}
for gauge-invariant vector variable $V_\mu$ we have
\begin{equation}\label{vector}
\left(\partial_z+3A^\prime\right)V_\mu=0,\quad \partial^\alpha\partial_\alpha V_\mu=0.
\end{equation}
Thus, similarly to abelian domain wall, from (\ref{graviton}) we find that the spectrum of tensor perturbations consists of a zero mode, or graviton, bound on the
brane, $\psi\sim e^{3A/2}$, and a set of continuous modes with $m^2>0$ move freely along the extra dimension. 
On the other hand, for the vector field  a normalizable solution for (\ref{vector}) is not feasible because $V_\mu(x,z)=e^{-3A(z)}V_\mu(x)$.


{\it Graviscalars}: in order to decouple the scalar sector the following constraints are required
\begin{equation}
 2\Gamma+\Theta=0,\quad  3A^\prime\Theta-3\Gamma^\prime-\phi_M^\prime\chi_M=0.\label{constrain}
\end{equation}
Thus, considering (\ref{constrain}) and the definition
\begin{equation}
e^{ip.x}\Omega(z)\equiv e^{3A/2}\Gamma/\phi_M^\prime ,
\end{equation}
we obtain
\begin{equation}\label{-ZZ}
\boldsymbol{Q}^+\boldsymbol{Q}\ \Omega(z)=m^2\Omega(z),
\end{equation}
with $\boldsymbol{Q}\equiv \partial_z+Z^\prime/Z$ and $\boldsymbol{Q}^+\equiv -\partial_z+Z^\prime/Z$, where
\begin{equation}
Z=e^{3A/2}\frac{\phi_M^\prime}{A^\prime}.
\end{equation}

Since the differential operator in (\ref{-ZZ}) is factorizable,  $m^2$ is real and positive and, hence, there are not unstable scalar excitations
in the spectrum of $\Omega$. On the order hand, as shown below, the massless graviscalar mode is not bounded around the brane \cite{Giovannini:2001fh}.

{\it Scalar perturbations}: similarly to the previous case, we define
\begin{equation}
e^{ip.x}\Xi_j(z)\equiv e^{3A/2}\left(\chi_j-\frac{\Gamma}{A^\prime}\phi_j^\prime\right),
\end{equation}
where, as noted above, the index $j$ indicates the component along the generator $\mathbf{T}^j$. In particular for $j=1$, associated with the direction along $\mathbf{M}$, the evolution equation for the gauge-invariant scalar fluctuations can be written as
\begin{equation}\label{Z-Z}
\boldsymbol{Q}\boldsymbol{Q}^+\ \Xi_M(z)=m^2\ \Xi_M(z).
\end{equation}
Notice that (\ref{-ZZ}) and (\ref{Z-Z}) can be viewed  as a SUSY quantum
mechanics problem \cite{ArkaniHamed:1999dc}. It follows that the eigenvalues of
$\Omega(z)$ and $\Xi_M(z)$ always come in pairs, except for the massless modes. Indeed,
\begin{equation}
\Omega(z)=\frac{1}{m}\boldsymbol{Q}^+\ \Xi_M(z)
\end{equation}
which exists strictly only for $m>0$. Moreover, the massless state
\begin{equation}
\Xi_M(z)\sim e^{3A/2}\frac{\phi_M^\prime}{A^\prime}
\end{equation}
is a non-normalizable mode and when gravity is switched off, this massless mode correspond to the bound translation mode of the flat space $SO(10)$ kink \cite{Shin:2003xy}. Then, in flat space there will exist the translation zero mode, which is removed from the four-dimensional KK spectrum when the extra dimension is warped \cite{George:2011tn}. For a single scalar field this conclusion was made in \cite{Shaposhnikov:2005hc}.

Now, it is not always the case that all zero modes of spin$-0$ fields are removed by the inclusion of warped gravity. For $j> 1$ we get
\begin{equation}\label{scalar2}
\left(-\delta_{j_1 j_2}\partial^2_z+V_{j_1 j_2}\right)\Xi_{j_2}(z)=m_{j_1 j_2}^2\ \Xi_{j_2}(z)
\end{equation}
with
\begin{equation}\label{Vq}
V_{j_1 j_2}=V_Q\delta_{j_1 j_2}+e^{2A}\frac{\partial^2V(\mathbf{\Phi})}{\partial\phi_{j_1}\partial\phi_{j_2}}\Bigg{|}_{\mathbf{\Phi}_k}
\end{equation}
where
\begin{eqnarray}\label{ddV}
\frac{\partial^2V}{\partial\phi_{j_1}\partial\phi_{j_2}}=-2\mu^2\delta^{j_1 j_2}
&&+3\lambda\text{Tr}[\mathbf{T}^{(j_1}\mathbf{T}^{j_3}\mathbf{T}^{j_4)}\mathbf{T}^{j_2}]\phi_{j_3}\phi_{j_4}\nonumber\\
+5\beta\text{Tr}[\mathbf{T}^{(j_1}\mathbf{T}^{j_3}\mathbf{T}^{j_4}&&\mathbf{T}^{j_5}\mathbf{T}^{j_6)}\mathbf{T}^{j_2}]\phi_{j_3}\phi_{j_4}\phi_{j_5}\phi_{j_6},
\end{eqnarray}
which is diagonal for each of the basis indicated in previous section, and $\mathbf{\Phi}_k$ is any non-abelian kink, $\mathbf{\Phi_A}$, $\mathbf{\Phi_B}$ or $\mathbf{\Phi_C}$, around which the perturbation is realized.

Next, let us will study the spectrum of eigenfunctions of (\ref{scalar2}) for $j>1$,  in correspondence with both the subgroup $H$ and the broken generators in the core of the wall.

\subsection{Symmetric kink}

For this scenario some components are subjected to the potential with
\begin{equation}\label{ddv1}
\frac{\partial^2 V}{\partial\phi_{j_1}^2}=-2k^2\left(1+\frac{2}{3}v^2-\left(3+\frac{2}{3}v^2(4-\frac{5}{3}F^2)\right)F^2\right)
\end{equation}
and others ones to the potential with
\begin{equation}\label{ddv2}
\frac{\partial^2 V}{\partial\phi_{j_2}^2}=-2k^2\left(1+\frac{2}{3}v^2(1 -\frac{1}{3}F^2)\right)(1-F^2).
\end{equation}
where $F\equiv\tanh(ky)$.

\begin{figure}[h]
\begin{center}
\includegraphics[width=8cm,angle=0]{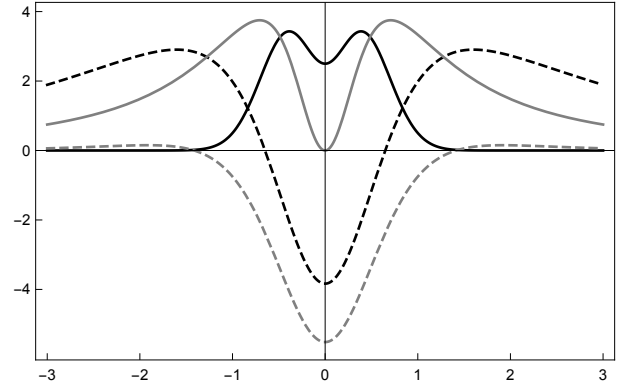}
 \caption{Plots of potential $V_j$ (dashed line) and the zero mode associated (solid line). Black line for (\ref{ddv1}) and gray line for (\ref{ddv2}).}
\label{VolcanoS}
\end{center}
\end{figure}

The plots depicted in Fig.\ref{VolcanoS} show that in both cases $V_j$ is a volcano potential. Notice that massive states have $m^2_j\geq 0$, where the zero modes for each component are bound states. Hence, there is no unstable tachyonic excitation in the system $\mathbf{\Phi_A}$.

On the other hand, the behavior of perturbations of the $SO(10)$ self-gravitating domain walls differs from the behavior of the exitations of the $SO(10)$ flat kinks where the $V_j$ are P\"oschl-Teller potentials \cite{Poschl:1933zz},
\begin{equation}\label{vqf1}
V_{j_1}=2\mu^2(3F^2-1), \qquad V_{j_2}=2\mu^2(F^2-1).
\end{equation}
For each spectrum of scalar states subjected to $V_{j_1}$ we find  two localized modes
\begin{eqnarray}
m_0^2=0,&&\quad \Xi_0\sim \cosh^{-2}(ky),\\
m_1^2=3\mu^2,\quad &&\Xi_1\sim \cosh^{-2}(ky)\sinh(ky) .
\end{eqnarray}
While for those ones under $V_{j_2}$ only a single state with negative eigenvalue is confined
\begin{equation}\label{Taquionic}
m_0^2=-\mu^2,\quad \Xi_0\sim \text{sech}(ky).
\end{equation}
This reveals the local instability of the symmetric kink when is embedded in a Minkowski spacetime.

When comparing with the $SO(10)$ warped scenario, we noticed that the gravitation repels the tachyonic mode and favors the four-dimensional localization of scalar states $\Xi_{j}(z)$, thus inducing the local stability of the scenario $\mathbf{\Phi_A}$.

\subsection{Asymmetric kink}

In Section \ref{solutions} we showed that on the domain wall $\mathbf{\Phi_B}$ the symmetry is broken from $SO(10)$ to the subgroups  $SO(6)$, $SU(2)$ and $U(1)$. In particular, along  the $SO(6)$ generators we find that the spectrum of scalar perturbations is restricted by $V_j$ which  depends on (\ref{ddv1}) or (\ref{ddv2}). In any case, a tower of states with positive eigenvalues is expected and hence  $\mathbf{\Phi_B}$ is perturbatively stable in these directions. 

\begin{figure}[h]
\begin{center}
\includegraphics[width=8cm,angle=0]{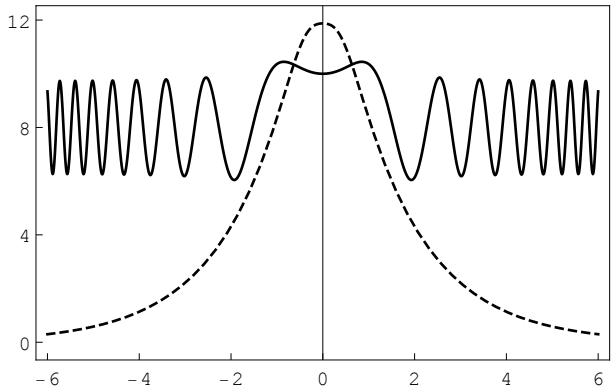}
 \caption{Plots of potential $V_{j_3}$ (dashed line) for the scalar perturbations (solid line) of $\mathbf{\Phi_B}$ along the $SU(2)\otimes U(1)$ generators.}
\label{barrera}
\end{center}
\end{figure}

On the other hand, for the components of $\Xi_{j}(z)$ along the generators of $SU(2)$ and $U(1)_{\mathbf{P_B}}$ we have
\begin{equation}\label{ddv3}
\frac{\partial^2 V}{\partial\phi_{j_3}^2}=4k^2\left(1+\frac{4}{9}v^2\right);
\end{equation}
so, $V_{j_3}$ is a positive barrier potential, see Fig.\ref{barrera}, which does not support  eigenfunctions with $m_{j_3}^2<0$. Therefore,  $\mathbf{\Phi_B}$ also is stable along the $SU(2)\otimes U(1)$ generators.

Now, when the gravity is switched off we find that the scalar perturbation hosted in $SU(2)\otimes U(1)$ are dominated by the potential $V_{j_3}=4\mu^2>0$ where the eigenvalues are defined for $m^2>0$. On the other hand, the wavefunctions associated to $SO(6)$ interact with the potentials (\ref{vqf1}) and once again within the spectrum of fluctuations there are tachyonic modes (\ref{Taquionic}) induced along the orthogonal subgroup. This puts in evidence the local instability of $\mathbf{\Phi_B}$ in five-dimensional Minkowski space.

\begin {figure}[h]
\begin{center}
\includegraphics[width=8cm,angle=0]{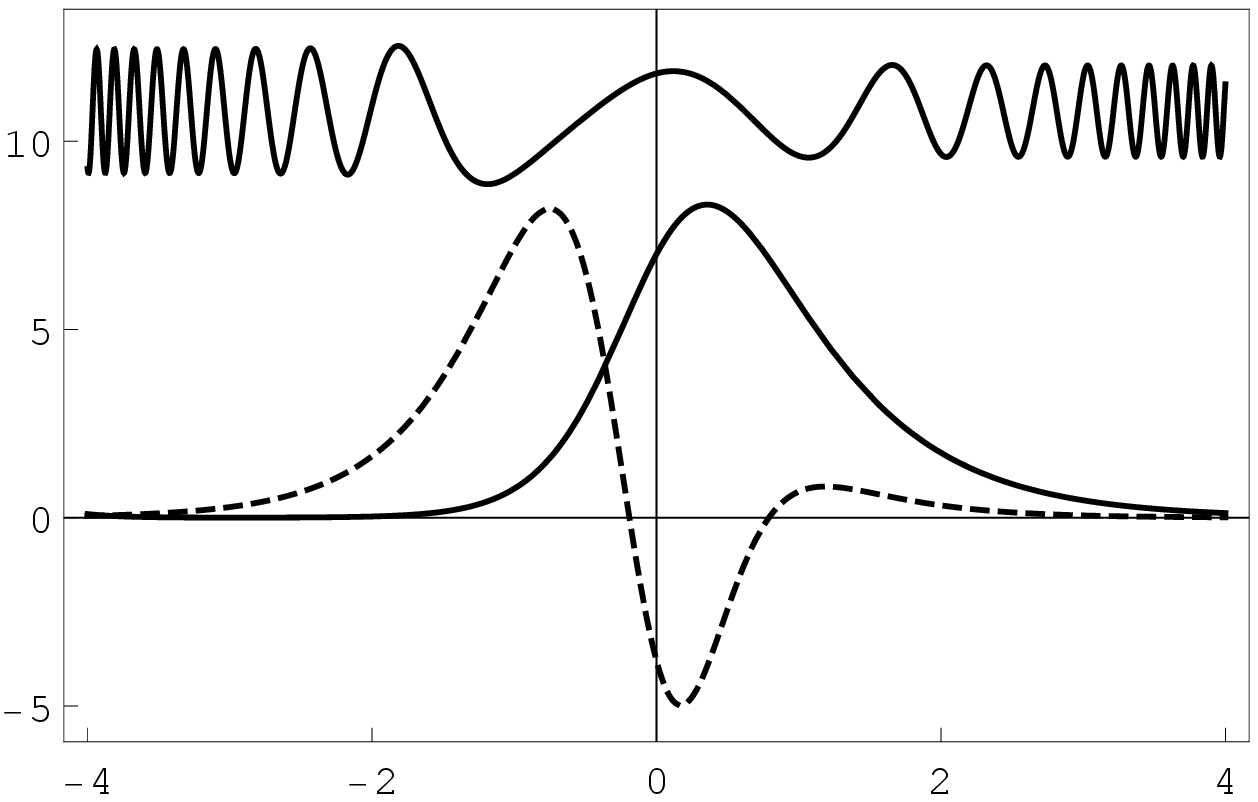}
\includegraphics[width=8cm,angle=0]{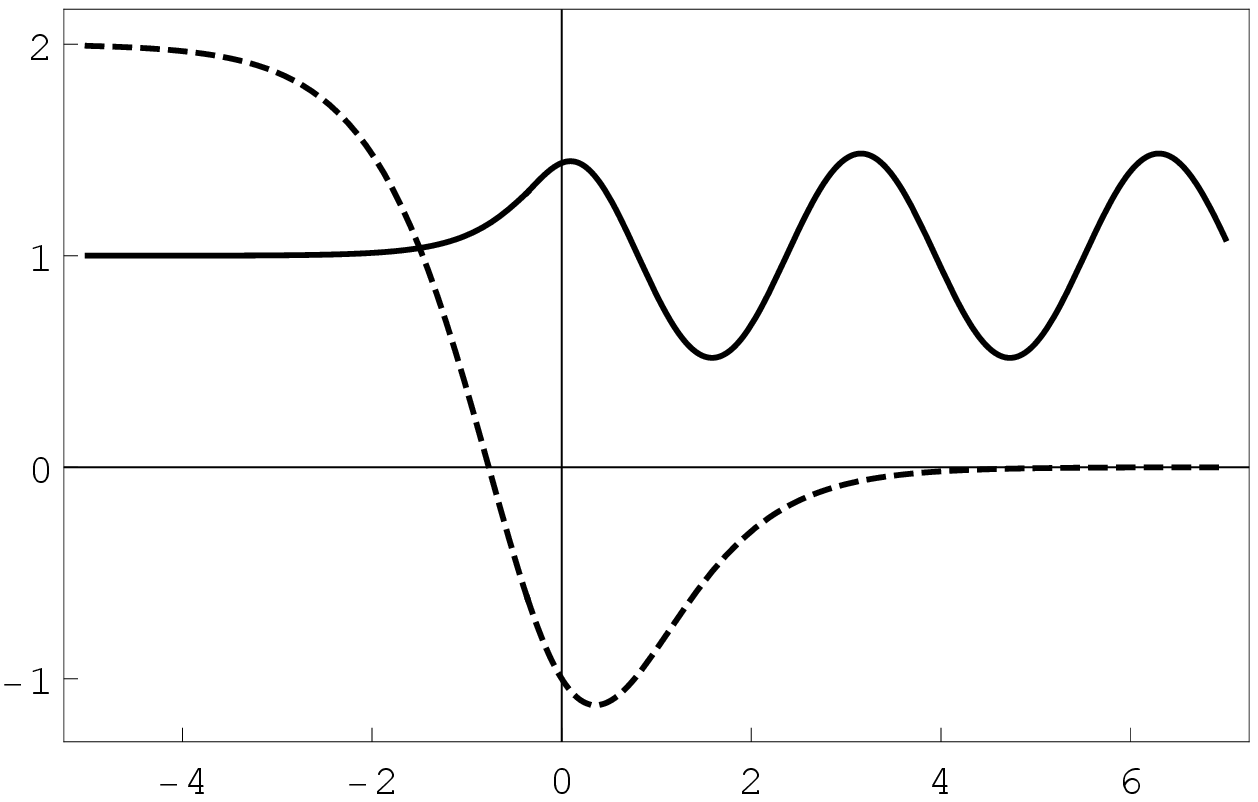}
\caption{Plot of the potentials $V_{j_-}$ (dashed line) and massive modes (solid line) for scalar perturbations of $\mathbf{\Phi_B}$ along the broken generator associated to $H_\mathbf{B}$, for  the warped geometry (top panel) and flat geometry (bottom panel).}
\label{VolcanoAF}
\end{center}
\end{figure}

With regard to the broken generators, for two components of scalar perturbation we find
\begin{equation}\label{ddv5}
\frac{\partial^2 V}{\partial\phi_{j_4}^2}=0
\end{equation}
which leads to a symmetric volcano potential for $V_{j_4}$ with $m_{j_4}\geq 0$ for the eigenfunctions associated. For the others twenty fourth fields we get
\begin{equation}\label{ddv4}
\frac{\partial^2 V}{\partial\phi_{j_\pm}^2}=2k^2\left(1+\frac{2}{3}v^2(1-\frac{1}{3}F^2)\right)(F\pm1)F
\end{equation}
and in this case, an asymmetric volcano potentials, $V_{j_\pm}$, is obtained. In Fig.\ref{VolcanoAF} (top panel)  the potential $V_{j_-}$ is shown (the profile of the potential $V_{j_+}$ is a specular image of the potential $V_{j_-}$; thus, both potentials have the same properties). The eigenfunctions are determined by a zero mode localized around the brane and a continuos tower of massive modes propagating freely for the five-dimensional bulk with $m_->0$. Additionally, due to the absence of $Z_2$ symmetry in the potential, resonance modes in the spectrum of fluctuations are expected to coexist \cite{Gabadadze:2006jm, Melfo:2010xu, Araujo:2011fm}. Hence, the perturbative stability of AdS${}_5$ vacua along the broken generators is guaranteed and $\mathbf{\Phi_B}$ is a stable braneworld.

Let us  comment a little further on the symmetry of the potential. For a single scalar field several asymmetric potentials arising from a spacetime without $Z_2$ symmetry have been found in \cite{Melfo:2002wd, Guerrero:2005aw}. However, in our case  the spacetime has $Z_2$ symmetry but not $V_{j_\pm}$. On the other hand, $\mathbf{\Phi}$ is a $SO(10)$ scalar field self-interacting  via $V(\mathbf{\Phi})$, i.e, the components $\phi_j$ of the field interact with each other according to the $SO(10)$ symmetry. Therefore, the $SO(10)$ group constrains on the self-interaction of field break the $Z_2$ symmetry of the scalar fluctuations along of the broken generators associated to $H_\mathbf{B}$.

Finally, in the flat scenario, where
\begin{equation}
V_{j_4}=0\label{ddv51} 
\end{equation}
and
\begin{equation}
V_{j_\pm}=2\mu^2(F\pm1)F,\label{ddv41}
\end{equation}
the last one plotted in Fig.\ref{VolcanoAF} (botton panel), we notice that the potentials do not support a normalizable zero mode. So, while the gravitation of the scenario delocalized the translation mode, it favours capture of others massless modes,  those ones along the broken generatos associated to $\mathbf{H_B}$. 

\subsection{Superasymmetric kink}
The scalar perturbation along $SO(2)$ is the translation mode and, according to what was shown at the beginning of this section, it is not located. In the directions of $SU(4)\otimes U(1)_{\mathbf{P_C}} $ we have the quantum mechanics potential (\ref{ddv3}) for the scalar perturbations. Hence, there are not normalizable massless states along  these generators. This also happens in flat case where $V_j=4\mu^2$ is obtained.

For the broken generators associated to $H_\mathbf{C}$, (\ref{ddv5}) is obtained for  twelve scalars and we get (\ref{ddv4}) for the last sixteen perturbations.  Thus, along the broken basis  massless bound states are found. Remarkably, in absence of gravity the analogous modes are not normalizable  since (\ref{ddv51}) and (\ref{ddv41})  are recovered \cite{Shin:2003xy}.

In any case we do not find  modes with $m^2 < 0$ and hence $\mathbf{\Phi_C}$ is stable under wall's perturbations.


 \section{Summary and Conclusions}\label{summary}

We have derived three $SO(10)$ self-gravitating kinks interpolating asymptotically between AdS${}_5$ vacuums, such that, whereas the symmetry breaking pattern $SO(10)\rightarrow SU(5)\times Z_2$ is  induced at the edges of the scenarios, in the core of each wall, a different unbroken symmetry is obtained: $SO(10)$, $SO(6)\otimes SU(2)\otimes U(1)/Z_2$ and $SU(4)\otimes SO(2)\otimes U(1)/Z_4$ respectively for the symmetric, asymmetric and superasymmetric kink.

These solutions are the gravitational analogue of the $SO(10)$ walls in Minkowskian  bulk found in \cite{Shin:2003xy}. The perturbative stability  of scenarios were studied and as a result we find that gravitation favors the stability of the $SO(10)$ walls and  its absence, on the contrary, weakens  the integrity of the scenarios.  In flat case, in addition to four-dimensional translation mode,  massive states and tachyonic P\"oschl-Teller states along $SO(10)$ and $SO(6)$ for the symmetric and asymmetric kink respectively are obtained in the spectrum of the scalar fluctuations. Fortunately, when gravity is included, the unstable tachyonic excitations are not already present and the scalar perturbation spectrum is defined only for $m^2\geq 0$.

The scalar fluctuations  of the non-abelian warped scenarios satisfy the following general characteristics: free massive modes ($m^2>0$), non-normalizable translation mode  and localized massless states along broken generators associated to $\mathbf{H_0}$ (Nambu-Goldstone bosons). These gravitational effects on the scalar fluctuations  are  fulfilled by the superasymmetric kink.

Now, for the symmetric and asymmetric kink, in addition to Nambu-Goldstone bosons, massless scalar excitations along the orthogonal subgroup are confined. The results are summered as follow: For the symmetric scenario, we find $SO(10)$ scalar zero modes  trapped by the wall. This effect  also is shared by the asymmetric scenario where scalar massless fluctuations along the  generators of $SO(6)$ are localized.  In both cases tachyonic modes are not found. Hence, the unstable modes  along the orthogonal groups  found in flat case are shifted for  bounded zero modes when gravity is included.

Finally, we observe that the interactions conditioned by the orthogonal symmetry, unlike  those ones defined by unitary group, could be favoring  the confinement of spinless bosons along the unbroken generators of $\mathbf{H_0}$. This issue is beyond the scope of this paper and will be treated in a next work.


\section{Acknowledgements}
 We wish to thank Rafael Torrealba and Oscar Castillo-Felisola for discussions.  This work was supported by CDCHT-UCLA under projects 007-RCT-2014 and 03-CT-2015.

\section{Appendix}\label{Appendix}

In $N$ dimensions one can define $N(N-1)/2$ linearly independent and antisymmetric matrices $\mathbf{L}$ to form a basis
such that any real antisymmetric $N\times N$ matrix can be expanded in terms of this basis. The Lie algebra for the basis $\mathbf{L}$ is given through
\begin{eqnarray}
[\mathbf{L}_{j_1 j_2},\mathbf{L}_{j_3 j_4}]=&&\delta_{j_1 j_4}\mathbf{L}_{j_2 j_3}-\delta_{j_1 j_3}\mathbf{L}_{j_2 j_4}\nonumber\\&&+\delta_{j_2 j_3}\mathbf{L}_{j_1 j_4}-\delta_{j_2 j_4}\mathbf{L}_{j_1 j_3},
\end{eqnarray}
where $j=1, \dots , N$. The mutually commuting generators can be found and they are $\mathbf{L}_{1 2}$, $\mathbf{L}_{3 4},\dots , \mathbf{L}_{N-1, N}$. These generators form an abelian subgroup i.e., the Cartan subalgebra of $SO(N)$. The rank of the algebra is equal to the number of mutually commuting generators.

A suitable generating expression for the basis $\mathbf{L}$ can be stated as
\begin{equation}
\left(\mathbf{L}_{j_1 j_2}\right)_{j_3 j_4}=\delta_{j_1 j_4}\delta_{j_2 j_3}-\delta_{j_1 j_3}\delta_{j_2 j_4}.
\end{equation}


In particular, for $N=10$ we deal with three kink solutions for the scalars field $\mathbf{\Phi}$  and to find explicitly the remain symmetry in the core of each kink we introduce three differently basis, A,B,C, obtained from a certain combination of ${\mathbf{L}}$'s.

{\it Basis A}: for the symmetric scenario
\begin{eqnarray}
\mathbf{T_A}^1&=&\mathbf{M_A},\\
\mathbf{T_A}^2=\frac{1}{\sqrt{20}}(\mathbf{L}_{34}+\mathbf{L}_{56}&+&\mathbf{L}_{78}+\mathbf{L}_{90}-4\mathbf{L}_{12}),\\
\mathbf{T_A}^3={\frac{1}{\sqrt{12}}}(\mathbf{L}_{56}+&\mathbf{L}_{78}&+\mathbf{L}_{90}-3\mathbf{L}_{34}),\\
\mathbf{T_A}^4=\frac{1}{\sqrt{6}}(\mathbf{L}_{78}&+&\mathbf{L}_{90}-2\mathbf{L}_{56}),\\
\mathbf{T_A}^5=\frac{1}{\sqrt{2}}(&\mathbf{L}_{90}&-2\mathbf{L}_{78}).
 \end{eqnarray}

{\it Basis B}: for the  asymmetric kink
\begin{eqnarray}
\mathbf{T_B}^1=\mathbf{M_B},\quad \mathbf{T_B}^{2}&=&\frac{1}{\sqrt{6}}(-2\mathbf{L}_{12}+\mathbf{L}_{34}+\mathbf{L}_{56}),\\
\mathbf{T_B}^{3}=\frac{1}{\sqrt{2}}(&\mathbf{L}_{34}&-\mathbf{L}_{56}),\quad\mathbf{T_B}^4=\mathbf{P_B}\\
\mathbf{T_B}^5&=&\frac{1}{\sqrt{2}}(\mathbf{L}_{78}-\mathbf{L}_{90}).
\end{eqnarray}

{\it Basis C}: for superasymmetric case
\begin{eqnarray}
\mathbf{T_C}^1=\mathbf{M_C}&,& \mathbf{T_C}^{2}=\mathbf{P_C},\\
\mathbf{T_C}^{3}=\frac{1}{\sqrt{12}}(-3\mathbf{L}_{34}&+&\mathbf{L}_{56}+\mathbf{L}_{78}+\mathbf{L}_{90}),\\
\mathbf{T_C}^4=\frac{1}{\sqrt{6}}(\mathbf{L}_{78}&+&\mathbf{L}_{90}-2\mathbf{L}_{56}),\\
\mathbf{T_C}^5=\frac{1}{\sqrt{2}}(&\mathbf{L}_{78}&-\mathbf{L}_{90}).
\end{eqnarray}

These  basis share forty generators which are determined by
\begin{equation}
\mathbf{T}^{j^\prime}=\frac{1}{\sqrt{2}}C^{j^\prime}_{i j}\mathbf{L}_{i j}, \quad {j^\prime}=6,\dots, 45,
\end{equation}
where $1/\sqrt{2}$ is a normalization factor and $C^{j^\prime}_{ij}$ a linear combination coefficient which is selected according to
\begin{equation}\nonumber
\begin{array}{c l l}
 & {j^\prime}=10+j,\quad & j\; \text{even}\\
C^{j^\prime}_{1j}=1,\quad & {j^\prime}=12+j,\quad & j\; \text{odd}\\
& {j^\prime}=3 + j,\quad & \text{for all}\; j;
\end{array}
\end{equation}
\vspace{0cm}
\begin{equation}\nonumber
\begin{array}{c l l}
& {j^\prime}=22+j,\quad & j\; \text{even}\\
C^{j^\prime}_{3j}=1,\quad & {j^\prime}=24+j,\quad & j\; \text{odd}\\
& {j^\prime}=17+j,\quad &  \text{for all}\; j;
\end{array}
\end{equation}
\vspace{0cm}
\begin{equation}\nonumber
\begin{array}{c l l}
& {j^\prime}=30+j,\quad & j\; \text{even}\\
C^{j^\prime}_{5j}=1,\quad & {j^\prime}=32+j, & j\; \text{odd}\\
& {j^\prime}=27+j,\quad &  \text{for all}\; j,
\end{array}
\end{equation}
for $10\geq j>i+1$; 
\begin{equation}\nonumber
\begin{array}{c}
C^{j^\prime}_{2j}=\left\lbrace\begin{array}{lll}
1,\quad&{j^\prime}=2+j,\quad & j\; \text{even}\\
&{j^\prime}=4+j,\quad & j\; \text{odd}\\
-1,\quad & {j^\prime}=11+j,\quad & \text{for all}\; j;
\end{array}
\right.
\end{array}
\end{equation}
\vspace{0cm}
\begin{equation}\nonumber
\begin{array}{c}
C^{j^\prime}_{4j}=\left\lbrace\begin{array}{lll}
1,\quad&{j^\prime}=16+j,\quad & j\; \text{even}\\
&{j^\prime}=18+j,\quad & j\; \text{odd}\\
-1,\quad & {j^\prime}=23+j,\quad & \text{for all}\; j;
\end{array}
\right.
\end{array}
\end{equation}
\vspace{0cm}
\begin{equation}\nonumber
\begin{array}{c}
C^{j^\prime}_{6j}=\left\lbrace\begin{array}{lll}
1,\quad&{j^\prime}=26+j,\quad & j\; \text{even}\\
&{j^\prime}=28+j,\quad & j\; \text{odd}\\
-1,\quad & {j^\prime}=31+j,\quad & \text{for all}\; j
\end{array}
\right.
\end{array}
\end{equation}
for $10\geq j>i$ and 
\begin{equation}
C^{42}_{70}=C^{45}_{79}=-1\nonumber
\end{equation}
\begin{equation}
C^{43}_{74}=C^{44}_{70}=C^{42}_{89}=C^{43}_{80}=C^{44}_{89}=C^{45}_{80}=1.\nonumber
\end{equation}

To indicate the unbroken symmetry group on the wall, we will focus on getting the basis that annihilate the field in the core,  $[{\mathbf{T}}, {\mathbf{\Phi}}(y=0)]=0$. For $\mathbf{\Phi_A}$ the result is straightforward because all generators annihilate to $\mathbf{\Phi_A}(y=0)$ and, therefore, the $SO(10)$ symmetry is restored on the kink.

For the asymmetric scenario $\mathbf{\Phi_B}$, there are nineteen generators annihilating the field in the origin of which fifteen of them form a basis for $SO(6)$ ($j=1$, 2, 3, 6, 7, 8, 9, 14, 15, 16, 17, 22, 23, 28, 29), three of them  ($j=5$, 42, 43) are generators of $SO(3)\sim SU(2)$ and the last one, $j=4$, in correspondence with $SO(2)\sim U(1)$. Hence, on the asymmetric kink $SO(10)\rightarrow SO(6)\otimes SU(2)\otimes U(1)/Z_2$ is obtained.

Finally, with respect to the superasymmetric kink $\mathbf{\Phi_C}$ we have seventeen generators annihilating the field in $y=0$. In this case, fifteen of them ($j=3$, 4, 5, 22, 24, 26, 28, 30, 32, 34, 36, 38, 40, 42, 43) are associated to $SO(6)\sim SU(4)$ and the two remaining ones ($j=1, 2$) are in correspondence with $SO(2)$ and with $SO(2)\sim U(1)$. Therefore, $SO(10)\rightarrow SU(4)\otimes SO(2)\otimes U(1)/Z_4$ is recovered in the core of the scenario.

Notice that, the unbroken symmetries $SO(6)\otimes SU(2)\otimes U(1)$ and $SU(4)\otimes SO(2)\otimes U(1)$ are closely related  with the Pati-Salam like group, $SU(4)\otimes SU(2)\otimes U(1)$, and the chiral bilepton gauge model, $SU(4)\otimes U(1)\otimes U(1)$, respectively.

\end{document}